# Extraordinary magnetooptical effects and transmission through the metal-dielectric plasmonic systems

V.I. Belotelov

General Physics Institute RAS, 38 Vavilov st., Moscow, 119991 Russia
M.V. Lomonosov Moscow State Univ., Moscow, 119992, Russia

L.L. Doskolovich

Image Processing Systems Institute RAS, 151, Molodog. st., Samara, 443001, Russia

A.K. Zvezdin

General Physics Institute RAS, 38 Vavilov st., Moscow, 119991 Russia

We report on significant enhancement of the magnetooptical effects in gyrotropic systems of a metallic film perforated by subwavelength hole arrays and a uniform dielectric film magnetized perpendicular to its plane. Calculations, based on a rigorous coupled-wave analysis, demonstrate the Faraday and Kerr effect spectra having several resonance peaks in the near infrared range, some of them coinciding with transmittance peaks. Qualitative analysis revealed that magnetic polaritons being coupled magnetic-film waveguiding modes with surface plasmons play a crucial role in the observed effect.

Corresponding author:
Dr. V.I. Belotelov
M.V. Lomonosov Moscow State University, Moscow, 119992, Russia
Tel.: +7-495-9391134
Fax : +7-495-9328820
Vladimir.belotelov@gmail.com

At present, magnetooptical (MO) Faraday and Kerr effects arising in gyrotropic media attract much attention due to their possible applications for control of light at a submicronic scale [1]. However, their values are not always sufficient, and the condition of getting simultaneous high transmission and Faraday rotation, which is of prime importance for optical devices, is often difficult to be fulfilled. In the last several decades the urge towards efficient MO media was mainly concentrated on the chemistry, i.e. on search for the optimal chemical composition of the magnetic substances [2]. Nevertheless, there are many constrictions in this approach, related to undesirably high absorption and Faraday ellipticity. At the same time, substantial MO effect enhancement can be achieved in some nanostructured materials owing to the materials' interesting optical properties inherent to their nanoscaled nature. This was shown recently by observation of giant Faraday effect in magnetic photonic crystals [3-5].

Giant Faraday effect can be achieved not only in photonic crystals but also in some other materials. In these respect metallic films perforated with the periodical arrays of subwavelength holes or groves can be of prime interest. Such systems have been extensively studied due to the phenomenon of the extraordinary optical transmission (EOT) found in them [6-7]. It was claimed in the most of the theoretical papers devoted to the EOT that the surface plasmon polaritons (SPPs) – collective oscillations of electron plasma at a metal surface coupled with photons (see e.g., Ref. [8]) – play a crucial role in the EOT. Though the precise involvement of the SPPs is still being debated, it is undoubted that they contribute in the EOT phenomenon allowing described structures to be referred as the plasmonic systems. Along with SPPs some other types of slow waves can be involved into the EOT, e.g. guided waves or surface waves supported by arrays of holes [9].

Magnetooptics of plasmonic structures has been considered only in several works so far [10,11]. In Ref. 9 optical transmission through a perforated metallic film in the presence of an external magnetic field applied in the film plane was studied and a strong dependence of the EOT-peak position on both the magnitude and direction of the in-plane magnetic field was found. Recently, experiments on perforated Co films have also been conducted [11]. In that study, the MO Kerr effect in the spectral range of the anomalous transmission band were found to be about one order smaller than that in uniform Co films of the same thickness.

In the present Letter we study MO effects in the bilayer heterostructure consisting of a periodically perforated nonmagnetic metallic plate deposited on a uniform magnetic dielectric layer (Fig. 1). We show that the Faraday and Kerr effects in this case can be enhanced significantly by appropriate choice of the geometry of the perforated heterostructure. It is important to note that the enhancement can be achieved in the vicinity of any desired wavelength from the range where efficient SPPs excitation in the metallic film is permitted.

To solve the problem of electromagnetic waves diffraction on the considered system a rigorous coupled-wave analysis (RCWA) was used [12,13]. In the RCWA technique the electromagnetic fields in each grating layer are determined by the modal approach. The electromagnetic boundary conditions are then applied at the interfaces between the substrate region, the individual grating slabs, and finally the superstrate region. The sequential application of electromagnetic boundary conditions reduces the computing of the reflected and the transmitted diffracted field amplitudes to the solution of linear system of equations. In our calculus we used a straightforward generalization of the RCWA method [12] onto the case of two-dimensional multilayered structures. In the modal approach to the computation of electromagnetic field in each grating layer we employed the correct rules of Fourier factorisation introduced in [13]. With the use of this method we obtain satisfactory convergence for multilayered structure containing crossed binary-relief binary grating made from highly reflecting metal. In numerical calculations 19x19 diffraction orders were retained (from order (-9, -9) to order (+9, +9)). To check convergence some results were also obtained with 31x31 orders. The results are in most cases the same within two decimal orders.

In the bilayer system we consider the magnetic layer uniformly magnetized in polar configuration, i.e. perpendicular to its plane. For the optical frequency range it is described by the dielectric tensor $\hat{\varepsilon}^{(m)}$ having the following nonzero components: $\varepsilon_{11} = \varepsilon_{22} = \varepsilon_{33} = \varepsilon_1$, $\varepsilon_{12} = -ig, \varepsilon_{21} = ig$, where $g$ is the value of medium's gyration [2]. Here we assume the case when magnetic medium is optically isotropic and second-order MO effects are negligible. In the numerical modeling parameters of Bi-substituted yttrium iron garnet magnetic film were used: $\varepsilon_1 = 5.5 + 0.0025i$, $g = (1 - 0.15i) \times 10^{-2}$ [2]. Metallic material is characterized by the dielectric function $\varepsilon_2$ with the Drude model $\varepsilon_2 = \varepsilon_\infty - \omega_p^2/(\omega^2 + i\gamma\omega)$ for the frequency dependence. At the analysis the metal film was assumed to be made of gold, for which we set $\varepsilon_\infty = 7.9$, $\omega_p = 8.77$ eV, and $\gamma = 1.13 \times 10^{14}$ c$^{-1}$ to fit empirical data [14] for the Au-film and the dielectric function $\varepsilon_2$ over wavelength range of interest from 750 nm to 1200 nm.

The results of the calculations of transmittance, reflectance and MO effects in the considered system are shown in Fig. 2. The Faraday and Kerr effects are described by the angles $\Phi_F$ and $\Phi_K$, which stand for the Faraday and Kerr rotation of light polarization, and angles $\Psi_F$ and $\Psi_K$ denoting Faraday and Kerr ellipticity of light polarization, respectively [2]. Transmittance spectrum of the Au/BiYIG bilayer has several EOT resonances, which are related to the light coupling with surface waves in the films. For the perforated metal films surrounded by two semi-infinite dielectric media the position of the transmittance peaks is well described by the condition of the SPPs excitation by the light of wavelength λ [6]:

$$\lambda_{max} = \frac{d}{\sqrt{u^2+v^2}} \left( \frac{\varepsilon_d \varepsilon_2}{\varepsilon_d + \varepsilon_2} \right)^{1/2} \qquad (1)$$

where $u$ and $v$ are integers determining SPPs modes, $\varepsilon_d = 1$ for Au/air interface and $\varepsilon_d = \varepsilon_1$ for Au/BiYIG interface. Strictly speaking, Eq. (1) is derived for the interface between uniform semi-infinite nonmagnetic media of metal and dielectric, but it is applicable for the qualitative analysis in the case considered here. Indeed, the correction to this formula related to the presence of magnetic field is shown to be of the next infinitesimal order [15]. In addition to that, it was shown in the previous studies that metallic film's perforation and finite thickness usually only slightly change the resonance condition [16]. For the considered case Eq. (1) for the near-infrared range predicts three resonances at λ$_{max}$=768 nm (SPP on Au/air interface), 871 nm, and 952 nm (SPP on Au/BiYIG interface). Direct calculations of the transmittance spectrum for the perforated metallic plate with dielectric layer of infinite thickness (semi-space) give very close values. However, it can be clearly seen from Fig. 2 (a) that when the perforated metal is adjacent to a dielectric slab of finite (submicron) thickness *h*, Eq. (1) is no longer appropriate to describe wavelengths for resonance transmittance. Consequently, some other types of surface waves become involved into the electromagnetic process and the problem of light interaction with such system gets much more complex.

The transmittance peaks are found to be strongly dependent on the thickness *h*, which is demonstrated by Fig. 3, where the transmittance and Faraday rotation versus the magnetic film thickness at fixed λ=971 nm are shown. It can be interpreted in terms of strong coupling between the incident field and waveguide eigenmodes supported by the dielectric slab. The SPPs should play a pronounced role in this coupling. Momentum of the incident light is divided between the SPP mode on the Au/BiYIG interface, hole arrays grating, and a waveguide wave into the BiYIG slab. Consequently, for the case of normal light incidence the conservation of momentum law gives

$$\beta \vec{e} = \vec{k}_{sp} \pm u_1 \vec{G}_x \pm v_1 \vec{G}_y ,\qquad(2)$$

where β is the propagation number of the waveguided mode, $\vec{e}$ is the unit vector along its propagation, $\vec{k}_{sp}$ is the SPP wave vector, $\vec{G}_x$ and $\vec{G}_y$ are the reciprocal lattice vectors for a square lattice with $|\vec{G}_x| = |\vec{G}_y| = 2\pi/d$, and $u_1$, $v_1$ are integers.

Thus, the dielectric layer attached to the perforated metal behaves as an optical planar waveguide, having one surface of dielectric medium and the other one of perforated metal. The condition for the excitation of a waveguide mode in the ray optics model of the wave propagation is written by:

$$2k_1 h \cos\theta + \Delta\varphi_1 + \Delta\varphi_2 = 2\pi m ,\qquad(3)$$

where $k_1 = 2\pi\sqrt{\varepsilon_1}/\lambda$ is the length of the wave vector in dielectric layer, $m$ is any integer, θ is an angle between the waveguide mode wave vector $\vec{k}_1$ and normal to the film's surface, and $\Delta\varphi_i$ are the phase shifts associated with the reflections at the dielectric/metal (*i*=1) and dielectric/air (*i*=2) boundaries (see Fig. 1). Phase shifts for the smooth metal surface $\Delta\varphi_1$ are found from the Fresnel formulae, but for the case of patterned metal their values deviate a little.

Simultaneous Eqs. (2) and (3) determine set of possible values of angle θ along with the dielectric film's thicknesses *h* for which a waveguide mode can be coupled with the SPPs for any fixed incident light wavelength. It should be noted here that due to the presence of hole arrays in the metal plate all waveguide modes scatter on the surface pattern and contribute into far-field intensity, i.e. into the observed transmittance. That is why it would be more correct to call these eigenmodes as quasi-guided modes. The calculations for the waveguide-mode supporting values of *h* at λ=971nm are indicated by vertical markers in Fig. 3 (solid lines for TE-mode and dotted lines for TM-mode). All transmittance peaks are very close to one of the resonance thickness value calculated from Eqs. (2-3), which proves the argument that the EOT in the considered systems is mainly due to the excitation of the quasi-guided modes in the dielectric plate. Higher deviations starting to appear for thicknesses larger than 350 nm can be attributed to the approximate nature of the given equations, which was accented in the previous paragraph.

Now let us discuss MO side of the problem. So far our reasoning has not taken into account the presence of the magnetization of the dielectric-plate. This is justified by the fact that medium's gyration is much smaller then the dielectric constant, and consequently can bring about relatively weak changes into the transmittance. At the same time, medium's gyrotropy should have significant influence onto the polarization of the transmitted or reflected light.

The latter is confirmed by results of simulation of Faraday and Kerr effects presented in Fig.2 (b-c). Pronounced MO effects enhancement is found at $\lambda_{max}$=971 nm, which is exactly the wavelength of the transmittance resonance. Namely, the Faraday and Kerr rotations at $\lambda_{max}$=971 nm get positive values of $\Phi_F = 0.74°$ and $\Phi_K = 0.35°$. This corresponds to the 8 and 5 times enhancement of the Faraday and Kerr rotation, respectively, in comparison with the same single magnetic layer surrounded by optically matched medium. Ellipticity of transmitted/reflected light polarization gets increased as well, but their positive and negative extrema happen at slightly different wavelengths. It is important to notice that the ellipticity is almost zero at the resonance of the MO rotation. Consequently, polarization states of transmitted and reflected linear polarized waves at λ=971 nm are very close to the linear polarizations.

It can be seen from Fig. 3, that peaks of the Faraday rotation and transmittance are related to each other and normally happen at close values of the thickness $h$. Faraday rotation has both positive and negative peaks, though for the single magnetic layer Faraday rotation is always negative (see dash-dotted line in Fig. 3). The analysis of the transmittance and Faraday effect dependences on the magnetic film's thickness $h$ reveals that for any wavelength of the incident light from the range where plasmons resonances exist it is possible to adjust the thickness to get significant Faraday rotation enhancement simultaneous with low light ellipticity and relatively high transmission.

In order to qualitatively explain obtained results one should take into account that in the presence of magnetization the dielectric plate becomes a MO waveguide, for which TM-mode – TE-mode conversion takes place. It should be emphasized here, that on the contrary to the conventional MO waveguides with in-plane magnetization [17], modes conversion for the considered structure is largely due to the leaking nature of the quasi-guided modes in the dielectric slab. When the thickness of the dielectric film is appropriate for the propagation of one of the quasi-waveguide mode coupled with SPP the polarization rotation arises and transmitted electromagnetic wave acquire significant Faraday angle. Let us mention here again, that waves in the dielectric slab scatter away, which is due to their interaction with the hole arrays pattern and to the MO modes conversion. It is interesting to note that the sign of the Faraday angle also depends on the type of the quasi-guided wave which supports the EOT at the given opto-geometrical parameters. Thus, the EOT peaks associated with TM-mode have the Faraday rotation of the same sign as for the single magnetic layer, while those peaks related to TE-mode have opposite Faraday rotation.

To demonstrate that SPPs are important for extraordinary MO effects we also modeled influence of SPPs damping on the transmittance and MO spectra. Characteristic parameter here is $\omega/\gamma$, where $\omega$ is the resonance frequency and $\gamma$ is the rate of electron collisions in the metal. It was found that when $\omega/\gamma \gg 1$ transmittance and the Faraday effect are high and resonance width is relatively small. However, when the parameter $\omega/\gamma$ approaches unity the EOT and Faraday rotation resonance peaks broaden, and get much smaller, completely vanishing for $\omega/\gamma \leq 1$, when the damping distance of the SPPs is about the lattice period. Such behaviour of extraordinary MO and transmittance spectra approves that the SPPs play a crucial role here.

To conclude, we have studied optical properties of bilayered system consisting of metal and dielectric parts, metallic plate being perforated with the periodic hole arrays and dielectric plate being magnetized in polar geometry. As test media for the metal and dielectric gold and BiYIG are used, respectively. Numerical modeling was performed on the basis of RCWA approach extended for the case of MO media. The effect of simultaneous enhancement of transmittance and MO Faraday and Kerr effect is found. It is shown that the proper choice of the gyrotropic nanostructure geometry allows acquiring the EOT and extraordinary Faraday rotation for any wavelength of the near-infrared range where the SPPs can be excited efficiently. Though the problem of light interaction with perforated metal-dielectric systems is very complex, qualitative explanations for the revealed phenomena are given, suggesting SPPs coupling with quasi-guided waves being in its roots. The effect of extraordinary Faraday rotation combined with the EOT is of practical interest taking into account numerous applications for the modern optical elements.

Work is supported by RFBR (04-01-96517, 05-02-17308, 05-02-17064, 06-02-17507), Russian President Grants (MK-3804.2005.2) and Russian foundation "Dynasty".

**Figures**

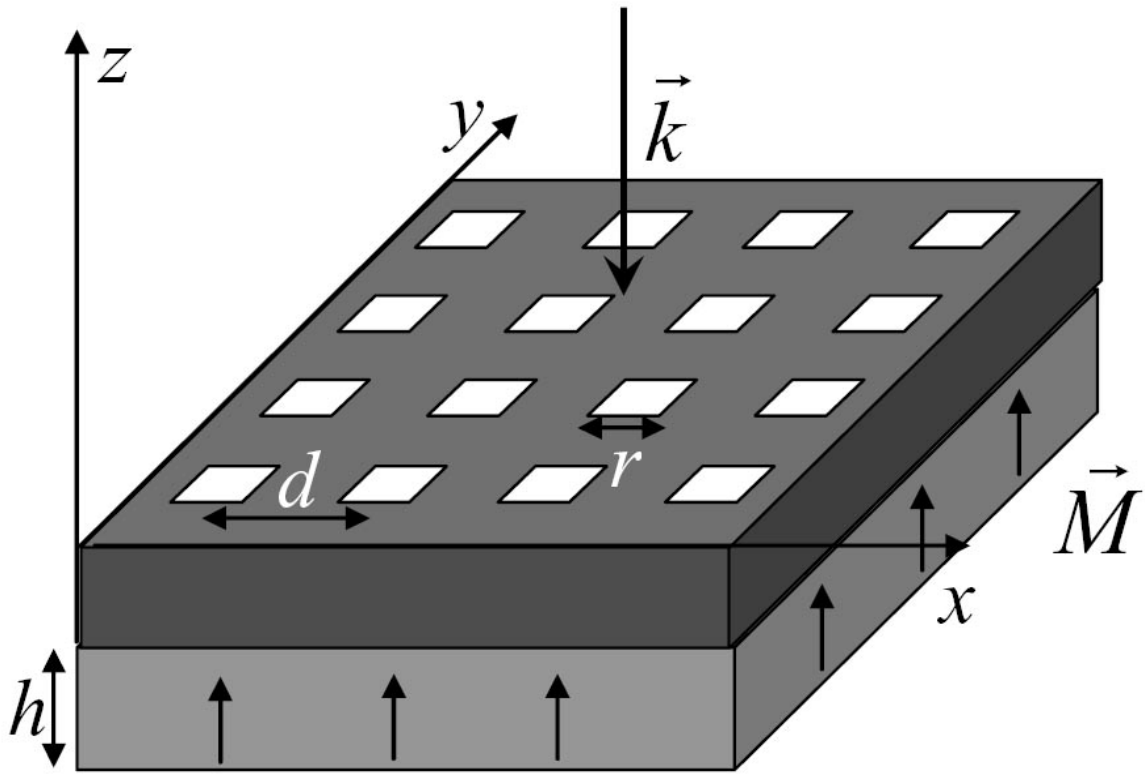

Figure 1. Problem configuration. Light with wave vector $\vec{k}$ is normally incident onto metal-dielectric bilayer. Metallic plate (upper layer) is periodically perforated with the square hole arrays. Holes constitute square lattice of period $d$. The size of each hole is $r$. Dielectric layer (lower layer) of thickness $h$ is magnetized in polar geometry (along OZ-axis).

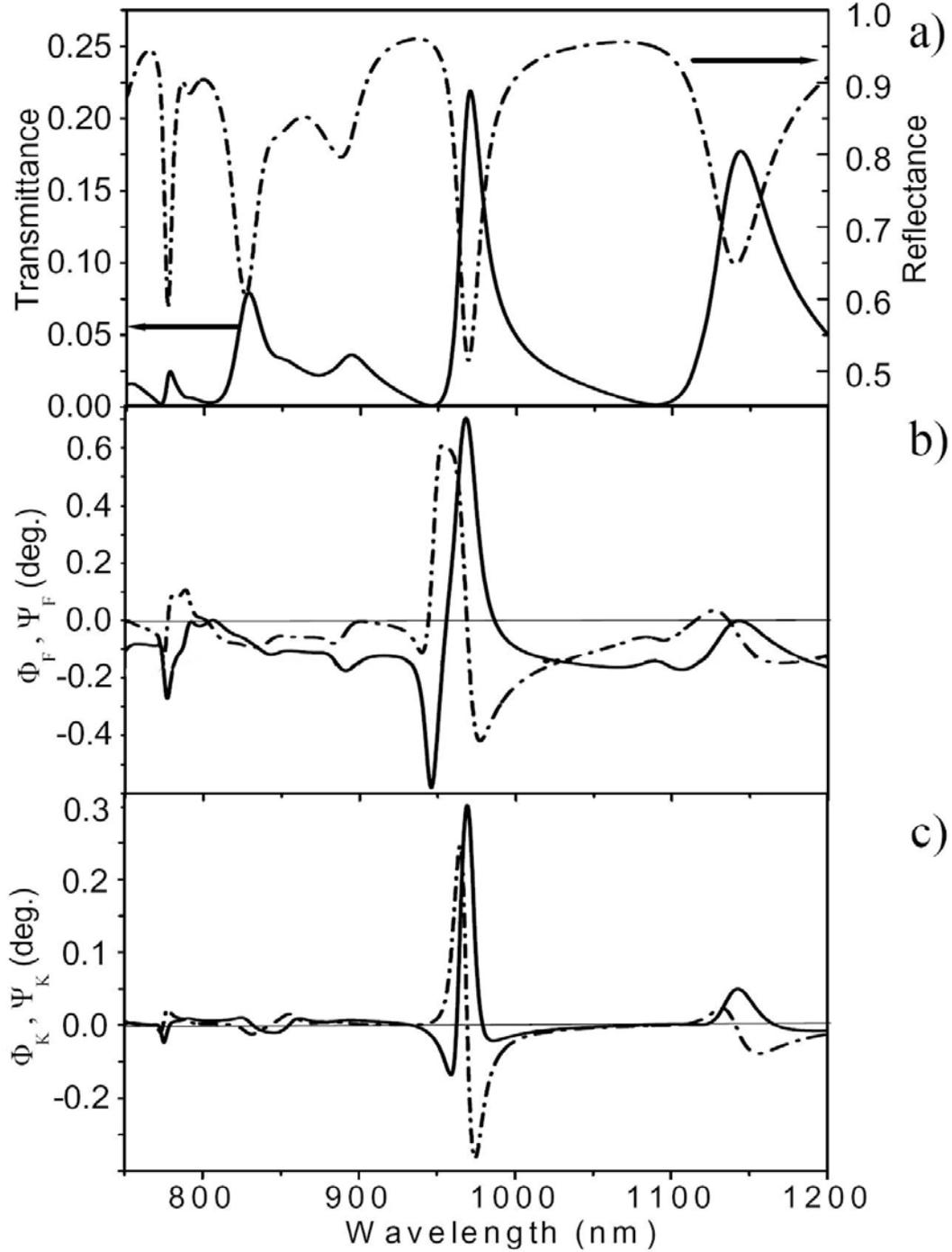

Figure 2. Spectra of optical transmittance (solid line on (a)), reflectance (dash-dotted line on (a)), Faraday rotation (solid line on (b)), Faraday ellipticity (dash-dotted line on (b)), Kerr rotation (solid line on (c)), Kerr ellipticity (dash-dotted line on (c)) of the bilayer system of perforated Au film of thickness 60 nm and uniform BiYIG film of thickness 120 nm. *d*=750 nm, *r*=260 nm (see Fig. 1).

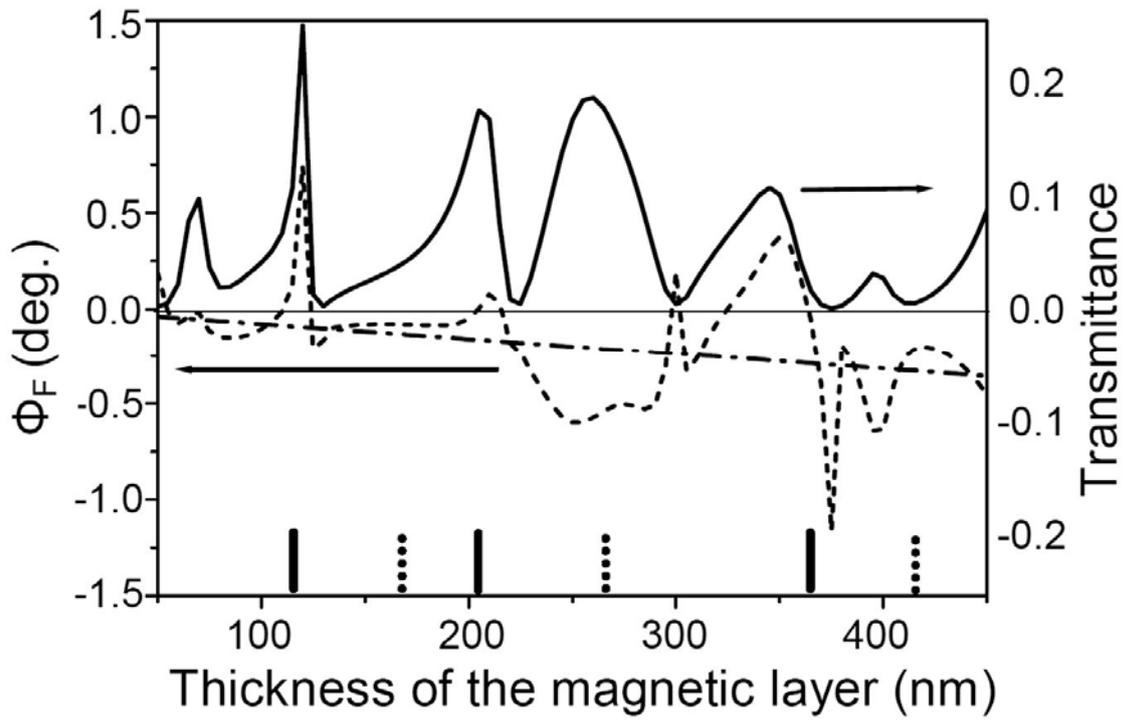

Figure 3. Faraday rotation and transmittance dependence on the magnetic layer thickness $h$ at the incident wavelength 971 nm. Geometrical parameters of the Au-BiYIG film are the same as in Fig. 2. Markers indicate the thicknesses for which dielectric slab supports either TE- (solid markers) or TM- (dotted markers) modes. Dashed-dotted line represents for comparison the Faraday rotation of single magnetic layer placed in optically matched surrounding medium.